\documentclass[iop]{emulateapj}

\shorttitle{A variable ULX in a globular cluster in NGC 4649}
\shortauthors{Roberts et al.}

\begin{document}


\title{A variable ultraluminous X-ray source in a globular cluster in NGC 4649}



\author{T.\,P. Roberts$^1$\email{t.p.roberts@durham.ac.uk}, G. Fabbiano$^2$, B. Luo$^2$, D.-W. Kim$^2$, J. Strader$^{2,3}$, M.\,J. Middleton$^1$, J.\,P. Brodie$^4$,\\ T. Fragos$^2$, J.\,S. Gallagher$^5$, V. Kalogera$^6$, A.\,R. King$^7$, A. Zezas$^8$}
\affil{
$^1$ Dept of Physics, Durham University, South Road, Durham DH1 3LE, United Kingdom\\
$^2$ Harvard-Smithsonian Center for Astrophysics, 60 Garden Street, Cambridge, MA 02138, USA\\
$^3$ Department of Physics \& Astronomy, Michigan State University, East Lansing, Michigan 48824, USA\\
$^4$ UCO/Lick Observatory, 1156 High St., Santa Cruz, Ca 95064, USA\\
$^5$ Department of Astronomy, University of Wisconsin, 475 N. Charter Street, Madison, WI 53706, USA\\
$^6$ Center for Interdisciplinary Exploration and Research in Astrophysics (CIERA) \& Department of Physics \& Astronomy,\\ Northwestern University, 2145 Sheridan Road, Evanston, IL 60208, USA\\
$^7$ Department of Physics \& Astronomy, University of Leicester, University Road, Leicester LE1 7RH, UK \\
$^8$ Physics Department, University of Crete, P.O. Box 2208, GR-710 03, Heraklion, Crete, Greece
}





\begin{abstract}
We report the discovery of a new ultraluminous X-ray source (ULX) associated with a globular cluster in the elliptical galaxy NGC 4649.  The X-ray source was initially detected with a luminosity below $5 \times 10^{38} \rm ~erg~s^{-1}$, but in subsequent observations 7 and 11 years later it had brightened substantially to $2 - 3 \times 10^{39} \rm ~erg~s^{-1}$.  Over the course of six separate observations it displayed significant spectral variability, in both continuum slope and absorption column.  Short-term variability in the X-ray flux was also present in at least one observation.  The properties of this object appear consistent with a stellar-mass black hole accreting at super-Eddington rates (i.e. in the ultraluminous accretion state), although a highly super-Eddington neutron star cannot be excluded.  The coincidence of an increase in absorption column with a possible enhancement in short-term variability in at least one observation is suggestive of a clumpy radiatively-driven wind crossing our line-of-sight to the object.

\end{abstract}





\section{Introduction}



Ultraluminous X-ray sources (ULXs, see \citealt{Roberts07}; \citealt{Feng11} for recent reviews) are most commonly found in star forming galaxies, with the link between the two phenomena spectacularly evident in the most actively star forming systems (e.g. \citealt{Fabbiano01}; \citealt{Gao03}).  However, not all ULXs are linked to star formation regions.  Surveys of ULXs have long identified a minority of them with the old stellar populations of elliptical galaxies (e.g. \citealt{Colbert02}; \citealt{Swartz04}), although a relatively large proportion of these ULXs may be mis-identifications of background objects, particularly when they have observed X-ray luminosities above $2 \times 10^{39} \rm ~erg~s^{-1}$ (\citealt{Irwin04}; \citealt{Walton11}).  Additionally, the recent discoveries of new ULXs in both M31 and M83 that have no young stellar counterpart indicates these objects must be low-mass X-ray binaries associated with an older stellar population (\citealt{Middleton12}; \citealt{Soria12}).

In recent years, a combination of the high X-ray spatial resolution and sensitivity of {\it Chandra\/} and {\it XMM-Newton\/} has begun to identify a population of ULXs in elliptical galaxies that have globular clusters (GCs) as optical counterparts.  These are particularly interesting given the surmise that GCs are a  plausible host for an intermediate-mass black hole (IMBH, e.g. \citealt{Miller04}).  Five good candidate GC-ULXs are now known: two in NGC 4472 (\citealt{Maccarone07}; \citealt{Maccarone11}), two in NGC 1399 (\citealt{Irwin10}; \citealt{Shih10}) and a new object in NGC 3379 \citep{Brassington12}.  Interestingly, optical spectroscopy of two of these objects shows remarkable emission line nebulae with strong [O III] lines and a lack of H lines (\citealt{Zepf08}; \citealt{Irwin10}); the origin of these nebulae remains a subject of much debate (e.g. \citealt{Maccarone11b}; \citealt{Ripamonti12}; \citealt{Clausen12}).

Here we report the discovery and characteristics of a new, variable ULX associated with a GC in the elliptical galaxy NGC 4649.  Although this object has been catalogued by various authors (\citealt{Randall04}; \citealt{Colbert04}; \citealt{Devi07}), this was from an April 2000 observation where it had a relatively low luminosity ($L_{\rm X} \sim 2 - 5 \times 10^{38} \rm ~erg~s^{-1}$, depending on assumed spectrum).  We report that in a sequence of observations taken in 2007 and 2011 it had brightened to the ultraluminous regime, and showed significant spectral and temporal variability.  In this paper we present the intriguing behaviour of this object, and discuss its nature.  We assume a distance of $d = 16.5$ Mpc to NGC 4649 in this work \citep{Blakeslee09}.


\begin{figure*}
\centering
\includegraphics[width=10cm]{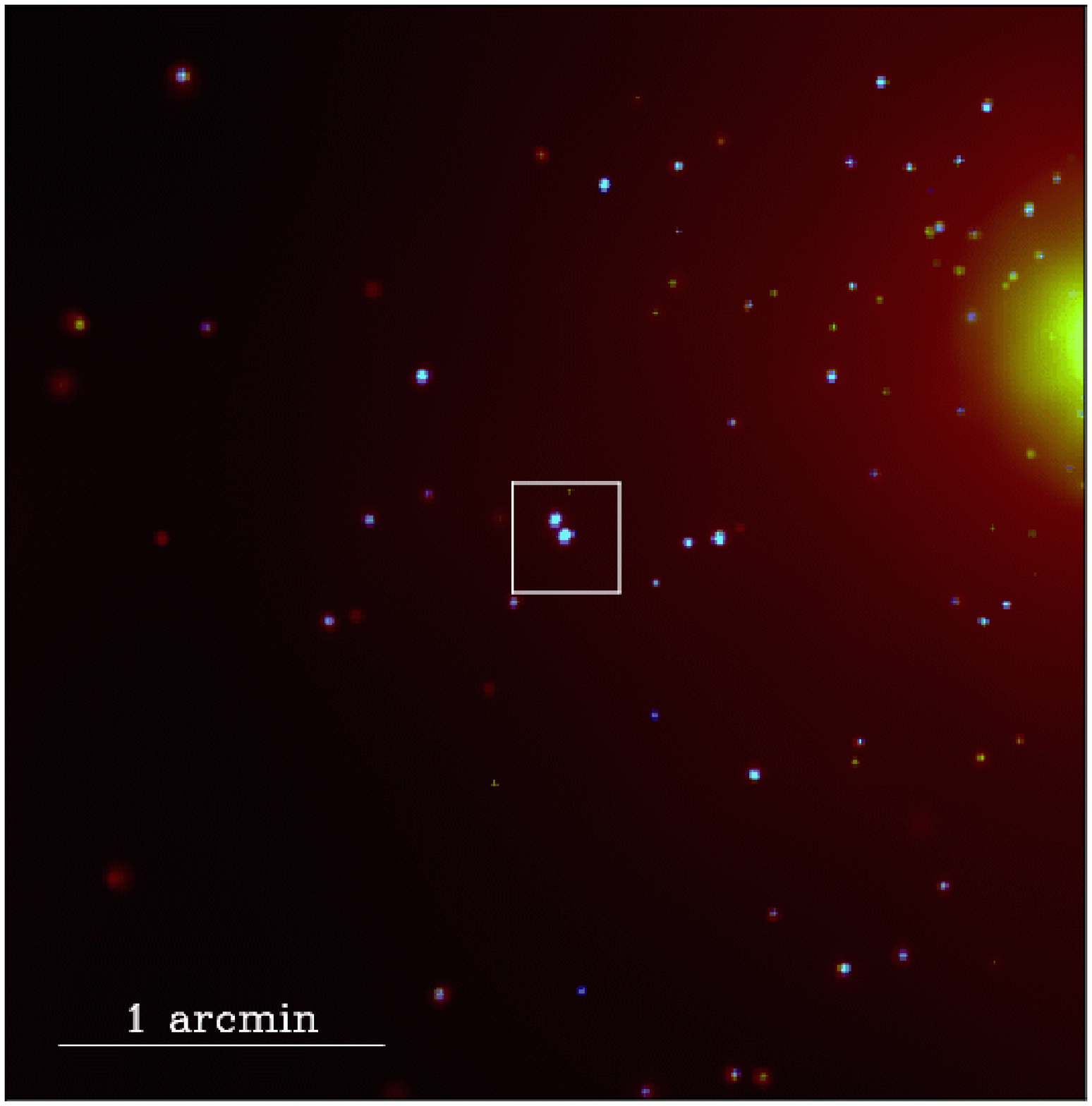}
\hspace*{1cm}
\includegraphics[width=4.5cm]{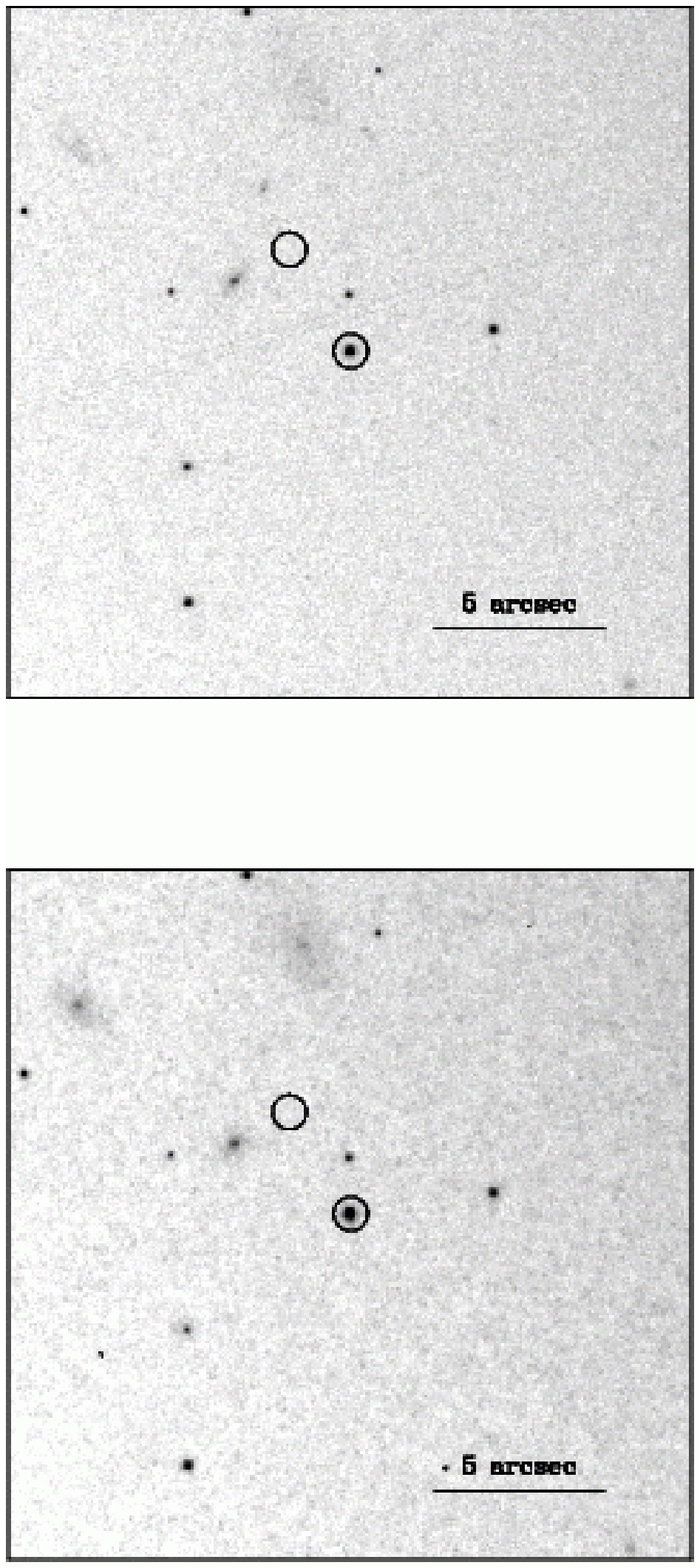}
\caption{{\it Left:\/} Three-colour {\it Chandra\/} ACIS-S X-ray image of the region containing CXOU J124346.9+113234.  It is visible in the centre of the image, close to the second ULX (see text).  The central regions of NGC 4649 are visible on the right of the image via their strong diffuse emission.  The square delimits the regions shown in the accompanying optical images.  The image uses data from the 0.3-1\, keV, 1-2\, keV and 2-8\, keV bands (shown as red, green and blue respectively), where each individual image is exposure-corrected and then adaptively smoothed using the CIAO tool {\sc csmooth}.  {\it Right:\/} {\it HST\/} ACS/WFC images of the region containing the ULXs, in the F475W ($\sim g$) and F850LP ($\sim z$) filters (upper and lower panels, respectively).  The positions of both ULXs are marked by open circles ({\bf not} scaled to the positional uncertainty).}
\label{ulximages}
\end{figure*}

\section{Data and source identification}

NGC 4649 has been observed six times by the {\it Chandra\/} ACIS-S detector, for a combined exposure of $\approx 300$ ks.  The details of these observations are summarised in Table~\ref{obsdets}.  The data were extracted from the {\it Chandra\/} data archive\footnote{{\tt http://cxc.harvard.edu/cda}}, and then cleaned, combined and searched for point-like sources using the standard {\sc ciao}\footnote{{\tt http://cxc.harvard.edu/ciao}} tools; the details of this analysis will be reported in \cite{Luo2012}.  Here we focus our attention on the brightest individual source detection, CXOU J124346.9+113234, that was detected with an average luminosity in excess of $2 \times 10^{39} ~\rm erg~s^{-1}$, placing it firmly in the ultraluminous regime.  As Table~\ref{obsdets} shows, a total of $\sim 1800$ counts were accumulated from the object, with in excess of 400 counts detected in each of the three longest observations.  

\begin{table}
\caption{Observation log}
\begin{center}
\begin{tabular}{lccc}\hline
ObsID	& Start time	& Exp. (ks)	& Counts \\
(1)	& (2)	& (3)	& (4) \\\hline
785 		& Apr 20 2000  3:34AM            &  34.2	& 67 \\
8182 	& Jan 30 2007 12:30PM           &  49.2	& 407 \\
8507 	&  Feb  1 2007  2:56AM            & 17.3	& 100 \\
12976 	&  Feb 24 2011  5:31PM           & 100.3	& 721\\
12975 	& Aug  8 2011  7:31AM            & 86.1	& 420 \\
14328 	&  Aug 12 2011  1:36AM           &  14.0	& 83 \\\hline
\end{tabular}
\end{center}
\begin{minipage}{8cm} Notes:  (1) {\it Chandra\/} observation identifier number.  (2) Observation start time and date (in UT).  (3) Observation exposure after cleaning for high background time intervals.  (4) Counts detected from each observation of the source, obtained from the X-ray spectrum.
\end{minipage}
\label{obsdets}
\end{table}

We show a portion of the ACIS-S field as a false-colour image in Fig.~\ref{ulximages}, centred on the position of the ULX.  Remarkably, given that only three ULX candidates were detected in NGC 4649, a second candidate ULX (CXOU J124347.0+113237 with $L_{\rm X} \approx 10^{39} \rm ~erg~s^{-1}$) lies less than 4 arcseconds from the position of our object.  This rules out using data from other missions with coarser spatial resolution to investigate the properties of either ULX; in particular two {\it XMM-Newton\/} observations totalling $> 145$ ks exposure are rendered useless given the $\sim 5$ arcsecond spatial resolution of its X-ray telescopes.  Happily, the {\it Chandra\/} detections are sufficiently separated to avoid cross-contamination of their data, and {\it Chandra\/} provides accurate positions that permit multi-wavelength searches for counterparts to both objects.  

\begin{figure*}
\centering
\includegraphics[angle=270,width=15cm]{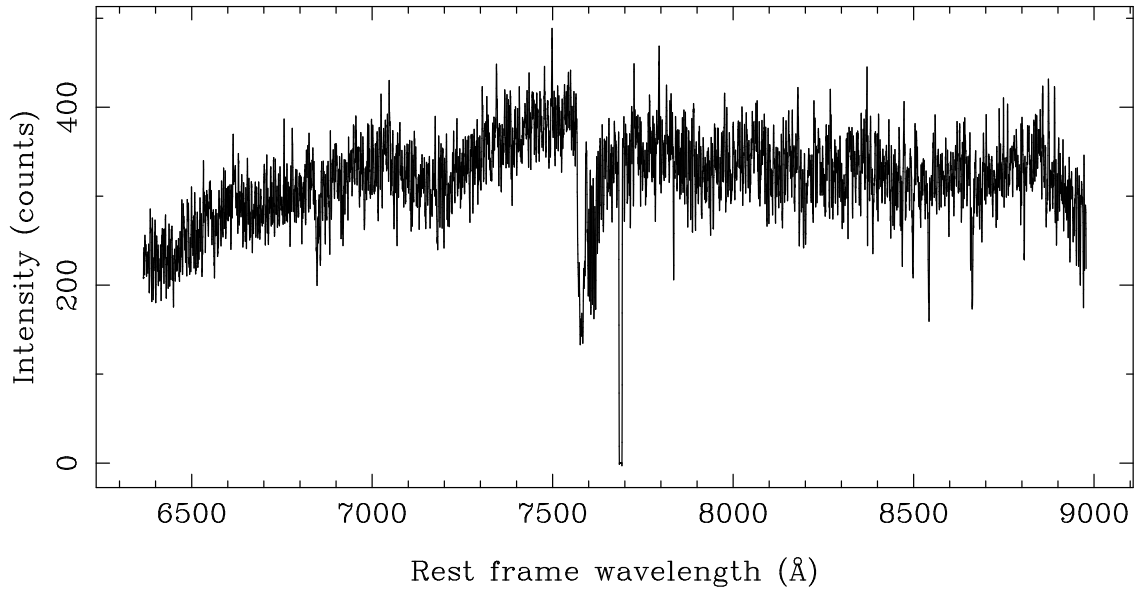}
\caption{Keck DEIMOS count spectrum of the optical counterpart to CXOU J124346.9+113234.  The spectrum is smoothed for display purposes by averaging over the five adjacent data points to each individual element, and corrected to its rest frame using the heliocentric velocity established from the Ca II triplet of absorption features at 8498, 8542 and 8662~\AA.  The other notable features include an instrumental chip gap at $\sim 7690$~\AA, the telluric Fraunhofer (A \& B) O$_2$ absorption features at $\sim 6850$ and $7580$~\AA\, (after correcting to the GC rest frame), and a H$\alpha$ absorption feature at $\sim 6563$~\AA.}
\label{optspec}
\end{figure*}

To the latter end, we have also obtained new {\it HST\/} ACS/WFC images in the F475W ($\equiv$ SDSS $g$) and F850LP ($\equiv$ SDSS $z$) filters.  We show a small region of each image, after preliminary cleaning and aligning to the X-ray data, in Fig.~\ref{ulximages}.  While no counterpart is observed for the fainter of the two ULXs, a clear counterpart is visible for CXOU J124346.9+113234, matching its position to $< 0.1$ arcseconds.  Further analysis (see \citealt{Strader12}) shows it to have $m_g = 21.81$ and $m_z = 20.26$ (accurate to $\pm 0.01$ magnitudes, and corrected for Galactic extinction as per \citealt{Peek10}), and half-light radii of $\sim 1.9$ pc.  These equate to absolute magnitudes $M_g = -9.28$ and $M_z = -10.83$ for NGC 4649, consistent with a GC.

However, a number of candidate ULXs discovered near to elliptical galaxies with bright optical counterparts have subsequently been identified as background QSOs (e.g. \citealt{Wong08}).  We therefore obtained an optical spectrum of the {\it HST\/} object with the DEIMOS spectrograph on the Keck II telescope. A 60 minute exposure was taken, using a 1200 line mm$^{-1}$ grating centred at 7800 \AA~and a 1 arcsecond slit, giving a resolution of $\sim 1.5$~\AA.  We show the spectrum in Figure~\ref{optspec}.  Through a cross-correlation of the region around the Ca II triplet of stellar absorption features we measure a heliocentric radial velocity of $1025\pm8$ km s$^{-1}$, consistent with the systemic velocity of NGC 4649 ($1117\pm6$ km s$^{-1}$; \citealt{Trager00}).  This provides strong evidence that the optical counterpart is indeed a GC associated with NGC 4649.  However, the spectrum shows no evidence for emission lines; in particular, H$\alpha$ is only seen in absorption, and the strong [NII] seen in the GC counterpart to CXOU J033831.8$-$352604 by \cite{Irwin10} is not observed. Unfortunately, the blue wavelength limit of the spectrum is $\sim 6400$\AA, so with these data we cannot determine whether there is any [OIII] emission as observed for the other two GCs with emission-line nebulae (\citealt{Zepf08}; \citealt{Irwin10}).





\section{X-ray characteristics}

Given that this is likely to be a GC-ULX in NGC 4649, its X-ray properties become of great interest.  We therefore extracted spectra and light curves for CXOU J124346.9+113234 from each observation.  Spectra were extracted using the standard {\sc ciao} tools from an aperture consistent with the 90\% encircled energy point spread function at 1.5 keV, and background spectra were obtained from the surrounding source-free regions.  For the three datasets with $> 400$ counts the spectra were grouped into bins of 20 counts or more to permit $\chi^2$ fitting in the X-ray spectral fitting package {\sc xspec}\footnote{\tt http://heasarc.nasa.gov/xanadu/xspec/}.  The remaining three datasets contain many fewer counts, and so were left unbinned and fitted to models in {\sc xspec} using the $C$-statistic.

\begin{figure*}
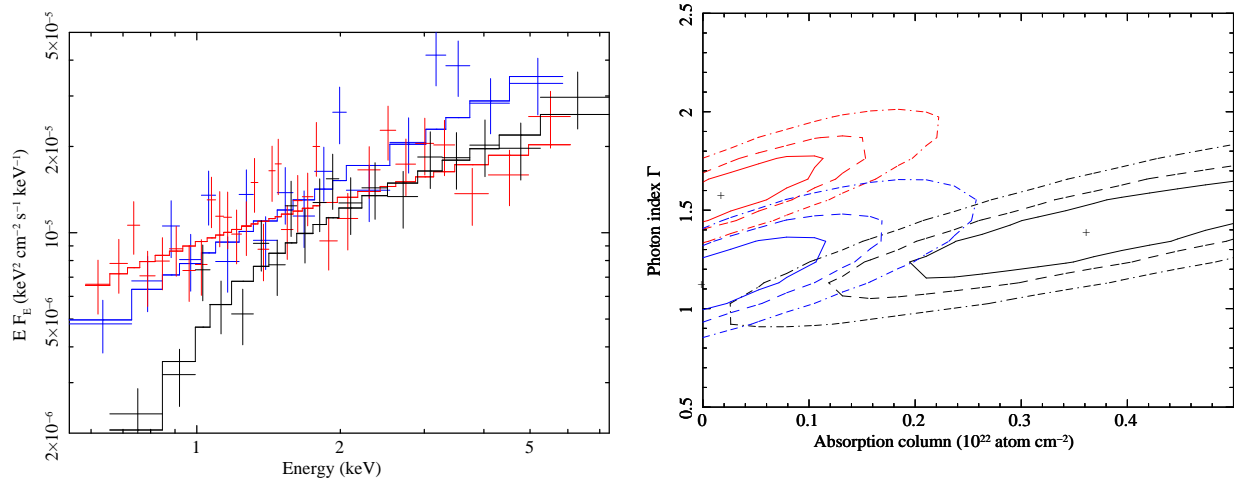

\centering
\includegraphics[angle=270,width=8cm]{fig3a.ps}
\includegraphics[angle=270,width=8.2cm]{fig3b.ps}
\caption{{\it Left:\/} X-ray spectra of CXOU J124346.9+113234, from the three best quality datasets.  The data and best-fitting absorbed power-law continuum models are unfolded from the instrument response, and each dataset is represented by a different colour: blue for ObsID 8182, red for 12976 and black for 12975.  {\it Right:\/} The differences in the spectra are emphasized by the confidence regions in the best fitting parameter values.  The 68\% (solid line), 95\% (dashed line) and 99\% (dot-dash line) confidence regions for the power-law photon index $\Gamma$ and the absorption column $N_{\rm H}$ are shown.  The colours are as per the left panel.}
\label{ulxspec}
\end{figure*}

We attempted only very simple model fits since the spectra were only moderate quality.  These were an absorbed power-law model ({\tt tbabs*po} in {\sc xspec}), and an absorbed multi-colour disc blackbody model (MCD; {\tt tbabs*diskbb}), consistent with the two dominant spectral models for Galactic black hole X-ray binaries (BHBs, cf. \citealt{McClintock06}).  In each case we added an additional fixed column of $2.2 \times 10^{20} \rm ~cm^{-2}$ to account for the absorption within our own Galaxy \citep{Dickey90}, and the absorption abundances were set as per \cite{Wilms00}.  The fitting was constrained to the 0.5-8\,keV band, where the detector response is best understood\footnote{We investigated the effect of extending these fits down to 0.3 keV, but for two of the three higher quality datasets no additional data were available below 0.5 keV.  The other spectrum (12976) had one additional data point; but this did not significantly change the best fitting parameters, and only modestly improved the error constraints (e.g. the 90\% constraint on the absorption column improved from $< 0.13$ to $< 0.12 \times 10^{22} \rm ~cm^{-2}$).}.  We detail the results of the fits in Table~\ref{specdets}.  The quoted errors on the best fitting parameter values are the 90\% confidence limits.  In the case of the data fitted using the $\chi^2$ method we constrained the flux by adding the multiplicative {\tt cflux} component to our models and re-fitting; for the poorer data we obtained the flux errors directly from the model normalisations.

\begin{table*}
\caption{Best fitting X-ray spectral parameters for CXOU J124346.9+113234}
\begin{center}
\begin{tabular}{lccccccc}\hline
ObsID	& \multicolumn{3}{c}{\tt tbabs*po}	&  \multicolumn{3}{c}{\tt tbabs*diskbb} \\
	&  $N_{\rm H}$	& $\Gamma$	& $\chi^2$/dof	& $N_{\rm H}$	& $kT_{\rm in}$	& $\chi^2$/dof	& $f_{\rm X,PL}$ \\
	& 	& 	& {\bf or} (C-Stat)	& 	& 	& {\bf or} (C-Stat)	& 	\\
(1)	& (2)	& (3)	& (4)	& (5)	& (6)	& (7)	& (8) \\\hline
785	& $< 0.20$	& $1.73^{+0.81}_{-0.41}$	& (170/509)		& $< 0.07$	& $0.93^{+0.59}_{-0.39}$	& (173/509)	& $1.09^{+0.71}_{-0.24}$\\
 \\
8182 & $< 0.13$	& $1.12^{+0.28}_{-0.15}$	& 19.43/15		& $< 0.04$	& $2.17^{+0.85}_{-0.49}$	& 22.43/15	& $8.96^{+1.23}_{-1.35}$\\
 \\
8507 & $< 0.47$	& $1.61^{+0.64}_{-0.52}$	& (242/509)		& $< 0.23$	& $1.29^{+0.69}_{-0.38}$	& (239/509)	& $5.05^{+4.04}_{-1.97}$\\
 \\
12976 	& $< 0.13$	& $1.58^{+0.23}_{-0.13}$	& 30.25/28	& $< 0.02$	& $1.25^{+0.18}_{-0.16}$	& 44.97/28	& $6.12^{+0.59}_{-0.67}$	\\
 \\
12975 	& $0.36^{+0.23}_{-0.20}$	& $1.39^{+0.29}_{-0.27}$	& 10.47/16		& $0.15^{+0.15}_{-0.13}$	& $2.16^{+0.93}_{-0.51}$	& 11.63/16	& $5.74^{+0.78}_{-0.76}$\\
 \\
14328 	& $< 0.38$	& $1.12^{+0.60}_{-0.39}$	& (229/509)		& $< 0.23$	& $2.18^{+3.24}_{-0.91}$	& (228/509)	& $6.81^{+5.70}_{-2.21}$\\\hline
 \end{tabular}
\end{center}
\begin{minipage}{18cm}Notes: (1) {\it Chandra\/} observation identifier number.  (2) \& (5) Absorption column external to our own Galaxy, in units of $10^{22} \rm ~atom~cm^{-2}$.  (3) Power-law photon index.  (4) \& (7) Goodness of fit statistic, either $\chi^2$/dof (where dof is the number of degrees of freedom) for the $\chi^2$ statistic, or C-stat/dof for the C-statistic, with the latter shown in parentheses for clarity.  (6) Inner disc temperature for the MCD model, in keV.  (8) Source flux, in the 0.5-8\,keV band and in units of $10^{-14} \rm ~erg~cm^{-2}~s^{-1}$, based on the power-law continuum fits.  The flux values for the MCD fits were generally 10 -- 20\% lower.
\end{minipage}
\label{specdets}
\end{table*}

The models provided acceptable fits to the data in all but one case, and even this was only marginally questionable (the MCD fit to ObsID 12976, with a null hypothesis probability of 0.022).  The parameters delivered by the fits vary in their ability to constrain the spectrum, with the poorer data yielding much larger errors.  Despite this, we are able to draw conclusions from the data.  It shows that the source spectrum is typically relatively unabsorbed, with most columns consistent with no absorption at the 90\% level, and most upper limits below a few $\times 10^{21} \rm ~cm^{-2}$.  The power-law continua are typically very hard, with indexes between 1.1 and 1.7, and the MCD fits appear hot ($kT_{\rm in} > 1$ keV) in most cases.  Crucially, the higher quality data demonstrates spectral variability.  We illustrate this in Fig.~\ref{ulxspec}, where we compare the three highest quality datasets.  The unfolded spectra from ObsIDs 8182 and 12976 in the left panel of Fig.~\ref{ulxspec} both appear unabsorbed, yet display markedly different slopes.  The 12975 data then appears different again, with strong absorption evident at low energies.  We emphasise these spectral differences by contouring the confidence regions for the power-law photon index and the absorption column in the right panel of the Figure.  None of the confidence regions overlap at the 95\% confidence level, and the ObsID 12976 and 12975 data are clearly different at the 99\% confidence level.  However, the spectral changes do not correlate directly with luminosity -- although the faintest and the most luminous epochs display the softest and (joint-)hardest spectra, these values are poorly constrained, and the intermediate luminosities show a range of behaviour that does not appear to scale with flux.

\begin{figure}
\includegraphics[angle=0,width=8.5cm]{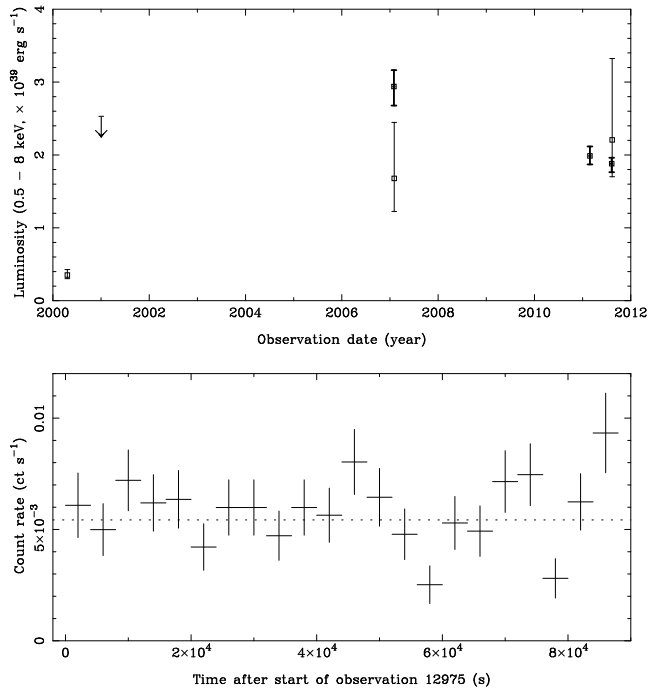}
\caption{{\it Top:\/} Long-term luminosity variations from CXOU J124346.9+113234.  The six observed 0.5-8\,keV luminosities from the {\it Chandra\/} ACIS-S observations are shown, with associated $1\sigma$ flux errors.  The additional $3\sigma$ upper limit arises from a 2XMM catalogue detection of the pair of ULXs (that are unresolvable by {\it XMM-Newton\/}).  The three better-quality datasets are highlighted by the thicker error bars.  {\it Bottom:\/} Light curve from the ObsID 12975 data, with the data in 4 ks bins.}
\label{ulxvar}
\end{figure}

As alluded to above, the ULX also displays long-term flux variability.  We show its long-term light curve in the top panel of Fig.~\ref{ulxvar}.  Clearly in 2000 the observation shows a source at slightly above the Eddington luminosity for a $1.4M_{\odot}$ neutron star; this subsequently rises to be persistently in the $2 - 3 \times 10^{39} \rm ~erg~s^{-1}$ regime, i.e. super-Eddington for a $10 M_{\odot}$ black hole, in the later observations.  It still varies strongly between the later observations -- for example, in 2007 the flux drops by a factor 2 in a day.

We also investigated the variability within each observation via two tests: the Gregory-Loredo (GL) algorithm \citep{Gregory92}, as implemented in the {\sc ciao} tool {\sc glvary}\footnote{\tt http://cxc.harvard.edu/ciao/threads/variable}; and a simple $\chi^2$ test against the hypothesis of a constant count rate.  We extracted light curves from a 1.75 arcsecond radius aperture centred on the ULX, with background again taken from the surrounding source-free regions.  The GL test reports variability in three observations -- ObsIDs 8507, 12975 and 14328 --  at a probability $> 0.9999$; the comparable probability for variability in the other three observations is $< 0.1$.  Unfortunately, the statistics for two of these observations are too poor to permit $\chi^2$ fitting of their light curve to confirm this variability, but we are able to do that for ObsID 12975, and this provides a marginal confirmation (null hypothesis probability of 0.012 that the light curve is constant; see also Fig.~\ref{ulxvar}).  We also confirm via $\chi^2$ fitting that the two other higher quality datasets (8182 \& 12976) show no evidence of variability.

\section{Discussion}

We have seen that CXOU J124346.9+113234 is a GC-ULX; that it displays long- and probable short-term flux variability; and that its spectrum also changes with time.  However, while we have been forced to fit the data with simple, single component models, Galactic BHBs are more commonly described by a combination of a power-law continuum and a MCD, plus absorption \citep{McClintock06}.  A scheme for interpreting the results of simple fits to the X-ray spectra of accreting binaries in elliptical galaxies (where the anticipated line-of-sight absorption in the host galaxy is minimal) in light of their likely greater complexity is presented by \cite{Brassington10}.  In their Fig. 16 they present a decision tree providing diagnoses of the underlying accretion state of an object, based on the absorption columns and goodness of fits to the same two simple models as we have used, only without an assumed foreground column.  We therefore re-fitted our spectral data, and used the results in combination with the decision tree to diagnose the behaviour of our ULX.  We provide a summary of this process, applied to our observations, in Table~\ref{Brassdiag}.

\begin{table*}
\caption{Diagnoses of source states according to the Brassington et al. (2010) scheme}
\begin{center}
\begin{tabular}{lcccc}\hline
ObsID	& $N_{\rm H, power-law}$	& MCD fit good?	& $N_{\rm H, MCD}$		& Diagnosis \\
(1)	& (2)	& (3)	& (4)	& (5) \\\hline
785		& $<$ Galactic	&	Marginal (goodness $\sim 91\%$)	& $= 0$	& steep power-law $+$ disc {\bf or} \\
& & & & cool disc $+$ non-thermal tail \\
8182		& $\sim$ Galactic	& Marginal (null hyp. $\sim 0.1$)	& $\approx 0$	& hard state {\bf or}\\
& & & & cool disc $+$ non-thermal tail \\
8507		&  $>$ Galactic		& Marginal (goodness $\sim 88\%$)	& $>$ Galactic	& source intrinsically absorbed {\bf or} \\
& & & & cool disc $+$ non-thermal tail \\
12976	& $\sim$ Galactic	& Marginal (null hyp. $\sim 0.04$)	& $= 0$	& cool disc $+$ non-thermal tail {\bf or} \\
& & & & hard state \\
12975	& $>$ Galactic		& Yes (null hyp. $\sim 0.7$)	& $>$ Galactic	& source intrinsically absorbed \\
14328	& $>$ Galactic		& Yes (goodness $\sim 56\%$)	& $>$ Galactic	& source intrinsically absorbed \\ \hline
\end{tabular}
\end{center}
\begin{minipage}{18cm}Notes: (1) {\it Chandra\/} observation identifier number.  (2) Absorption column fitted as part of a simple {\tt tbabs*po} model, compared to the Galactic foreground column of $2.2 \times 10^{20} \rm ~cm^{-2}$.  (3) Does a {\tt tbabs*diskbb} model constitute an acceptable fit to the data?  Here, we use a null hypothesis of $< 0.05$ (for the $\chi^2$ statistic) or goodness of $> 95\%$ (C-stat)  as the criteria for an unacceptable fit, hence for the four datasets close to this cut-off we provide alternative diagnoses of the state.  (4) Absorption column for the {\tt tbabs*diskbb} model.  (5) State diagnosis.
\end{minipage}
\label{Brassdiag}
\end{table*}

Interestingly, its behaviour appeared to split into two groups.  The first (ObsIDs 785, 8182 \& 12976) could all be interpreted as a cool MCD spectrum ($kT_{\rm in} < 0.5$ keV) plus a non-thermal component, although given the uncertainties on the fits a power-law dominated hard state or a steep power-law with either a cool or hot ($> 1.5$ keV) disc, were also possible interpretations.  Crucially, in the absence of very high quality data, a cool disc plus non-thermal tail spectrum is consistent with the ultraluminous state spectrum as described by \cite{Gladstone09}, indicative of super-Eddington emission from stellar-mass BHBs.  The second group (8507, 12975 \& 14328) 
were all best described by an absorbed spectrum, in which the underlying model cannot be determined, although the same cool disc plus non-thermal tail model was a possible alternative interpretation for observation 8507.  Interestingly, these three observations were the same three observations in which the GL test detected variability.  This classification is supported by the spectral fits in Table~\ref{specdets} -- 12975 is the only observation in which absorption is clearly detected, and (unlike the other three datasets) the best-fitting column was well above Galactic ($\sim 10^{21} \rm ~cm^{-2}$) for 8507 and 14328, although their poor data quality mitigated against excluding null columns.

To put our object in context, it is instructive to consider the other members of this class.  The primary importance of these objects was as the first positive evidence for black holes in GCs, when previously there was considerable uncertainty that GCs could host such objects \citep{Maccarone07}\footnote{The recent report of two black hole candidates in the Galactic GC M22, detected as flat-spectrum radio sources, now provides further evidence that black holes might be present in some numbers in GCs; see \cite{Strader12a}.}.  Given the evidence for short- and long-term variability in our object, and the high luminosities it attains, CXOU J124346.9+113234 is therefore another good candidate black hole in a GC.  The evidence so far points to GC-ULXs being stellar-mass black holes accreting at super-Eddington rates.  This is indicated firstly by the possible presence of massive outflows from these objects (where massive, radiatively-driven outflows are a key prediction of super-Eddington accretion, see e.g. \citealt{Poutanen07}) , seen in optical \citep{Zepf08} and X-ray \citep{Brassington12} spectroscopy.  Secondly, the intrinsic X-ray spectra of the objects themselves betray signs of super-Eddington accretion.  For example, the spectra of both the object in RZ 2109, and CXOKMZJ033831.7$-$353058 in NGC 1399, are cool-disc-dominated when in the ultraluminous regime (\citealt{Maccarone07}; \citealt{Shih10}), consistent with an ultraluminous state spectrum where the soft component (either the outer disc, or perhaps more plausibly the photosphere at the base of a radiatively-driven outflow) dominates, cf. \cite{Gladstone09} for examples.  Furthermore, the NGC 1399 object fades in flux with time and possesses a classic thermal-dominant spectrum at sub-Eddington fluxes, consistent with a stellar-mass black hole.  On the other hand, the second GC-ULX in NGC 4472 possesses a hot disc-like spectrum when ultraluminous \citep{Maccarone11}; but this is also consistent with many ULXs as they reach super-Eddington fluxes, e.g. \cite{Middleton11}.

Perhaps most pertinently the GC-ULX in NGC 3379 has an underlying hard, power-law continuum spectrum \citep{Brassington12}, similar to our object; and the variability seen in the RZ 2109 object on both short and long timescales can be attributed to variations in the absorption column close to the object \citep{Maccarone11}.  So, although we cannot definitively rule out the presence of an IMBH in CXOU J124346.9+113234, a more likely scenario (based on the \citealt{Brassington10} diagnoses) is that we are observing a stellar-mass object in the ultraluminous state.  We therefore speculate that the ULX, when first observed, was in a steep power-law state with a significant disc contribution; it then brightened significantly and in the 2007 and 2011 observations it was in an ultraluminous state.  In the latter observations there were indications of increased absorption column when the flux diminished, consistent with the behaviour seen in the RZ 2109 GC-ULX.  The most important new evidence from our object is the possible link between enhanced variability and absorption, primarily in ObsID 12975 but also very tentatively in both 8507 and 14328.  This is plausible evidence for an outflow driven by super-Eddington processes, if the increase in variability is the result of a clumpy radiatively-driven wind passing across the line-of-sight to the X-ray emitting regions of the ULX, and so simultaneously both providing additional absorption and an extrinsic source of variability (cf. \citealt{Middleton11b} for the luminous ULX NGC 5408 X-1).  We therefore conclude that our object has properties consistent with a stellar-mass black hole accreting in the super-Eddington ultraluminous state. 

Interestingly, if it is showing distinctly super-Eddington behaviour at only $2-3 \times 10^{39} \rm ~erg~s^{-1}$ then this argues that the compact object is relatively small, likely $< 10 M_{\odot}$.  This is consistent with BH formation in a metal-enriched environment (see below), where the remnants are not expected to be very massive (e.g. \citealt{Fryer12}).  Indeed, given the peak luminosity of $\sim 3 \times 10^{39} \rm ~erg~s^{-1}$, a highly super-Eddington neutron star cannot be entirely excluded as a plausible explanation for this ULX.  To this end we note that the Z sources often appear at super-Eddington luminosities, with Circinus X-1 in particular having been seen to reach up to 10 times its Eddington limit, and to show a soft, disc-dominated spectrum similar to high accretion rate black holes at such fluxes \citep{Done03}\footnote{After some uncertainty, the neutron star nature of Circinus X-1 has recently been confirmed by renewed bursting activity (see \citealt{D'ai12} and references therein); however its distance remains the subject of debate, and the peak luminosity may be a up to factor 0.3 times lower.}.  A pertinent scenario for a neutron star hosted in a GC was suggested by \cite{King11}, who notes that the brightest X-ray sources in GCs could be mildly-beamed super-Eddington neutron stars in ultra-compact X-ray binaries; however difficulties with this scenario were discussed by \cite{Peacock12}.  Clearly the current data is unable to unambiguously distinguish whether a black hole or neutron star underlies CXOU J124346.9+113234; unfortunately such a definitive resolution (e.g. through the detection of type-I bursts) appears unlikely for any GC-ULX with the current generation of instrumentation.

Finally, \cite{Maccarone11} note that the host clusters for the previously discovered GC black hole candidates are typically redder (and hence more metal rich; although see \citealt{Brassington12}) and more luminous (and hence more massive) than an average GC in their host galaxies.  The GC host to CXOU J124346.9+113234 fits into this pattern; its $g-z$ colour and $z$ magnitude are in the top $\sim 20\%$ reddest/brightest GCs in NGC 4649, and its mass is estimated at $2.8 \times 10^6 M_{\odot}$ \citep{Strader12}.  In addition, its relatively small half-light radius is also consistent with the notion that X-ray binaries are more readily formed in clusters with higher collision rates (e.g. \citealt{Peacock10}).  This fairly typical habitat for an X-ray binary argues against any excessively exotic interpretation for the object; once again a super-Eddington X-ray binary would appear a plausible interpretation.  

\section{Conclusions}

We have reported the detection of a new ULX coincident with a GC in the galaxy NGC 4649.  As with other members of the still-rare class of GC-ULXs, this object is a good candidate to be a black hole radiating at super-Eddington luminosities, although we cannot entirely exclude the possibility that this is a highly super-Eddington neutron star.  The benefit of having had a series of {\it Chandra\/} observations spaced over 11 years is that it has allowed us to demonstrate that this source varies its characteristics with time; the \cite{Brassington10} simulations show that the observed spectra can be interpreted as super-Eddington emission, with some observations showing an additional absorption column.  The observations with additional column also show at least some evidence for short-timescale variability; together this could be interpreted as evidence for a clumpy, radiatively driven wind (a key prediction of super-Eddington accretion) crossing the line-of-sight, consistent with some other ULXs.  Interestingly, this again points to a commonality in physical processes between ULXs in both old and young stellar populations (cf. \citealt{Middleton12}).  CXOU J124346.9+113234 therefore serves to highlight the potentially interesting astrophysics that we can recover from the less well-studied ULXs associated with older stellar populations, both tentatively with current missions and, potentially, in far more detail with future high collecting area and high spatial resolution X-ray observatories.




\acknowledgments

The authors thank an anonymous referee for comments that have improved this paper.  TPR thanks the Royal Society for the award of an International Exchange scheme grant.  This work was supported under NASA grants GO-12369.01-A HST (PI Fabbiano) and GO1-12110X Chandra (PI Fabbiano).  We acknowledge support from the CXC, which is operated by the Smithsonian Astrophysical Observatory (SAO) for and on behalf of NASA under contract NAS8-03060.  GF thanks the Aspen Center for Physics.



{\it Facilities:} \facility{Keck}, \facility{CXO (ACIS)}.


\begin{thebibliography}{}
\bibitem[Blakeslee et al.(2009)]{Blakeslee09} Blakeslee J.P., et al., 2009, ApJ, 694, 556
\bibitem[Brassington et al.(2010)]{Brassington10} Brassington N.J., et al., 2010, ApJ, 725, 1805
\bibitem[Brassington et al.(2012)]{Brassington12} Brassington N.J., et al., 2012, ApJ, 755, 162
\bibitem[Clausen et al.(2012)]{Clausen12} Clausen D., Sigurdsson S., Eracleous M., Irwin J.A., 2012, MNRAS, 424, 1268
\bibitem[Colbert \& Ptak(2002)]{Colbert02} Colbert E.J.M., Ptak A.F., 2002, ApJS, 143, 25
\bibitem[Colbert et al.(2004)]{Colbert04} Colbert E.J.M., Heckman T.M., Ptak A.F., Strickland D.K., Weaver K.A., 2004, ApJ, 602, 231
\bibitem[D'ai et al.(2012)]{D'ai12} D'ai A., et al., 2012, A\&A, 543, A20
\bibitem[Devi et al.(2007)]{Devi07} Devi A.S., Misra R., Agrawal V.K., Singh K.Y., 2007, ApJ, 664, 458
\bibitem[Dickey \& Lockman(1990)]{Dickey90} Dickey J.M., Lockman F.J., 1990, ApJ, ARA\&A, 28, 251
\bibitem[Done \& Gierli{\'n}ski(2003)]{Done03} Done C., Gierlinski M., 2003, MNRAS, 342, 1041
\bibitem[Fabbiano, Zezas \& Murray(2001)]{Fabbiano01} Fabbiano G., Zezas A., Murray S.S., 2001, ApJ, 554, 1035
\bibitem[Feng \& Soria(2011)]{Feng11} Feng H., Soria R., 2011, New AR, 55, 166
\bibitem[Fryer et al.(2012)]{Fryer12} 	Fryer C.L., Belczynski K., Wiktorowicz G., Dominik M., Kalogera V., Holz D.E., 2012, ApJ, 749, 91
\bibitem[Gao et al.(2003)]{Gao03} Gao Y., Wang Q.D., Appleton P.N., Lucas R.A., 2003, ApJ, 596, L171
\bibitem[Gladstone, Roberts \& Done(2009)]{Gladstone09} Gladstone J.C., Roberts T. P., Done C., 2009, MNRAS, 397, 1836
\bibitem[Gregory \& Loredo(1992)]{Gregory92} Gregory P.C., Loredo T.J., 1992, ApJ, 398, 146 
\bibitem[Irwin et al.(2010)]{Irwin10} Irwin J.A., Brink T.G., Bregman J.N., Roberts T.P., 2010, ApJ, 712, L1
\bibitem[Irwin, Bregman \& Athey(2004)]{Irwin04} Irwin J.A., Bregman J.N., Athey A.E., 2004, ApJ, 601, L143
\bibitem[King(2011)]{King11} King, A., 2011, ApJ, 732, L28
\bibitem[Luo et al.(2012)]{Luo2012} Luo B., et al., 2012, {\it in prep.}
\bibitem[McClintock \& Remillard(2006)]{McClintock06} McClintock J.E., Remillard R.A., 2006, in Lewin W.H.G., van der Klis M., eds, Compact Stellar X-ray Sources. Cambridge Univ. Press, Cambridge, p. 157
\bibitem[Maccarone et al.(2007)]{Maccarone07} Maccarone T.J., Kundu A., Zepf S.E., Rhode K.L., 2007, Nature, 445, 183
\bibitem[Maccarone et al.(2010)]{Maccarone10} Maccarone T.J., Kundu A., Zepf S.E., Rhode K.L., 2010, MNRAS, 409, L84
\bibitem[Maccarone et al.(2011)]{Maccarone11} Maccarone T.J., Kundu A., Zepf S.E., Rhode K.L., 2011, MNRAS, 410, 1655
\bibitem[Maccarone \& Warner(2011)]{Maccarone11b} Maccarone T.J., Warner B., 2011, MNRAS, 410, L32
\bibitem[Middleton et al.(2011)]{Middleton11b} Middleton M., Roberts T.P., Done C., Jackson F., 2011, MNRAS, 411, 644
\bibitem[Middleton, Sutton \& Roberts(2011)]{Middleton11} Middleton M.J., Sutton A.D., Roberts T.P., 2011, MNRAS, 417, 464
\bibitem[Middleton et al.(2012)]{Middleton12} Middleton M.J., Sutton A.D., Roberts T.P., Jackson F.E., Done C., 2012, MNRAS, 420, 2969
\bibitem[Miller \& Colbert(2004)]{Miller04} Miller M.C., Colbert E.J.M., 2004, IJMPD, 13, 1
\bibitem[Peacock et al.(2010)]{Peacock10} Peacock M.B., Maccarone T.J., Kundu A., Zepf S.E., 2010, MNRAS, 407, 2611
\bibitem[Peacock, Zepf \& Maccarone(2012)]{Peacock12} Peacock M.B., Zepf S.E., Maccarone T.J., 2012, ApJ, 752, 90
\bibitem[Peek \& Graves(2010)]{Peek10} Peek J.E.G., Graves G.J., 2010, ApJ, 719, 415
\bibitem[Poutanen et al.(2007)]{Poutanen07} Poutanen J., Lipunova G., Fabrika S., Butkevich A.G., Abolmasov P., 2007, MNRAS, 377, 1187
\bibitem[Randall, Sarazin \& Irwin(2004)]{Randall04} Randall S.W., Sarazin C.L., Irwin J.A., 2004, ApJ, 600, 729
\bibitem[Roberts(2007)]{Roberts07} Roberts, T.\,P.,  2007, \apss, 311, 203
\bibitem[Ripamonti \& Mapelli(2012)]{Ripamonti12} Ripamonti E., Mapelli M., 2012, MNRAS, 423, 1144
\bibitem[Shih et al.(2010)]{Shih10} Shih I.C., Kundu A., Maccarone T.J., Zepf S.E., Joseph T.D., 2010, ApJ, 721, 323
\bibitem[Soria et al.(2012)]{Soria12} Soria R., Kuntz K.D., Winkler P., Blair W.P., Long K.S., Plucinsky P.P., Whitmore B.C., 2012, ApJ, 750, 152
\bibitem[Strader et al.(2012a)]{Strader12a} Strader J., Chomiuk L., Maccarone T.J., Miller-Jones J.C.A., Seth A.C., 2012a, Nature, 490, 71
\bibitem[Strader et al.(2012b)]{Strader12} Strader J., et al., 2012b, ApJ, {\it in press} (arXiv:1210.3621v1)
\bibitem[Swartz et al.(2004)]{Swartz04} Swartz D.A., Ghosh K.K., Tennant A.F., Wu K., 2004, ApJS, 154, 519
\bibitem[Trager et al.(2000)]{Trager00} Trager S.C., Faber S.M., Worthey G., Gonz{\'a}lez J.J., 2000, AJ, 119, 1645
\bibitem[Walton et al.(2011)]{Walton11} Walton D.J., Roberts T.P., Mateos S., Heard V., 2011, MNRAS, 416, 1844
\bibitem[Wilms, Allen \& McCray(2000)]{Wilms00} Wilms J., Allen A., McCray R., 2000, ApJ, 542, 914
\bibitem[Wong, Chornock \& Filippenko(2008)]{Wong08} Wong D.S., Chornock R., Filippenko A.V., 2008, PASP, 120, 266
\bibitem[Zepf et al.(2008)]{Zepf08} Zepf S.E., et al., 2008, ApJ, 683, L139

\end{thebibliography}
\end{document}